# Proposition d'une méthode de qualification et de sélection d'un logiciel d'analyse et de suivi du référencement dans les moteurs de recherche

**Sébastien Bruyère [1,2], Vincent Pillet [2], Luc Quoniam [1]**
[1] **Université du Sud Toulon Var**
[2] **BleuRoy.com, Marseille, France**

**ABSTRACT**. In order to measure website visibility in search engines, there are softwares for analytics and referencing follow-up. They permit to quantify website's efficacity of referencing and optimize its positionning in search engines. With regard to search engines' algorithms' evolution and centralization of Key Performance Indicators for Marketing decision making, it becomes hard to find solutions to effectively lead a lot of projects for referencing. That's why we have built a methodology in order compare, evaluate and choose a software for analytics and referencing follow-up in search engines.

**Introduction**

Alors que les prévisions affichent que 69% des annonceurs prévoient de maintenir l'utilisation du référencement naturel pour la promotion d'un site Web en 2009 (Benchmark Groupe, 2009), il apparait essentiel d'évaluer l'efficacité du référencement naturel et des actions d'optimisation associées. Mais devant la multitude de solutions disponibles sur le marché, il devient difficile d'identifier une solution cohérente aux objectifs des entreprises dans le domaine.

Dans ce cadre, nous avons élaboré une méthodologie pour qualifier et sélectionner rapidement des logiciels, puis, nous l'avons appliquée pour désigner et choisir un logiciel d'analyse et de suivi de positionnement dans les moteurs de recherche le plus adapté à nos besoins.

## 1 Matériels & Méthodes

Contrairement au cas des logiciels libres[1], il existe aucune démarche vulgarisée permettant de qualifier et de sélectionner un logiciel propriétaire sans faire appel aux méthodes de mathématique appliquées pour l'aide à la décision. Ces dernières, bien que très performantes, sont souvent lourdes et difficiles à mettre en œuvre dans une entreprise qui doit rapidement faire des choix.

Par conséquent nous avons opté pour revenir à des méthodes plus traditionnelles basées sur la mobilisation de l'intelligence collective dans le cadre de réunions interactives.

Ainsi, nous avons organisé une première « réunion de réflexion » ([Zar04]) avec l'ensemble des acteurs en charge des activités de référencement, puis au moyen de la méthode du « brainstorming » (Osborn, 1953), nous les avons interrogés sur les critères prépondérants quant à l'analyse et au suivi de référencement dans les moteurs de recherche pour nos clients.

Nous avons retenu la méthode de brainstorming car « *elle favorise l'échange, l'inter-connaissance, elle permet de mettre au clair les conceptions de chacun dès le départ, évitant les risques de conflits ultérieurs* » ([TD02]).

Par ailleurs l'organisation sous forme de « réunion de réflexion » a permis de provoquer « *des discussions que chacun peut avoir sur les idées des autres. Chaque groupe ayant une vision particulière liée à un niveau hiérarchique, à un métier, une expertise, une expérience, une ancienneté dans l'entreprise.*

*La réflexion collective ne peut déboucher que si l'on confronte les réponses des différents groupes par paliers successifs. Chaque palier va aider à formuler de nouvelles réponses liées à l'agrégation des groupes, dans des groupes de taille supérieure jusqu'à n'obtenir qu'un seul groupe.* » ([Zar04]).

Partant de ces approches, nous avons constitué un comité de pilotage composé ainsi:
- Le Directeur Technique et R&D : Il dispose d'une vision panoramique sur les aspects techniques et sur l'application des recherches

---

[1] Plusieurs méthodes existent : OSSM CapGemini (Capgemini, 2003), OSSM Navica (Navicasoft, 2004), QSOS (Atos Origin, 2004) et OpenBRR (Carnegie Mellon West, SpikeSource, O'Reilly, Intel, 2005)

réalisées. Il est intervenu sur les aspects d'interconnexion entre les différentes solutions.
- Le consultant e-Marketing : En contact permanent avec la clientèle, il a pu positionner certains critères émanant des attentes clients.
- Les WebMarketeurs : Ils réalisent des rapports de positionnement au quotidien, leurs apports sur les aspects de pilotage à la réalisation ont été importants.
- Le Chef de Projet R&D : En veille permanente sur les nouvelles technologies, il a pu intervenir sur les aspects innovants et les nouveaux indicateurs clés de performance pour le référencement.

A partir de la liste des différents critères proposés, nous avons recherché un outil capable d'établir une comparaison qualitative collective entre différents logiciels.

Nous avons opté pour une matrice de pondération logiciels/critères à évaluer collectivement.

|           | **Logiciel 1**    | **Logiciel 2**    | **Logiciel 3**    |
|-----------|-------------------|-------------------|-------------------|
| **Critère 1** | Evaluation 1 à 3 | Evaluation 1 à 3 | Evaluation 1 à 3 |
| **Critère 2** | Evaluation 1 à 3 | Evaluation 1 à 3 | Evaluation 1 à 3 |
| **Critère 3** | Evaluation 1 à 3 | Evaluation 1 à 3 | Evaluation 1 à 3 |

**Fig. 1** *Matrice de pondération (exemple)*

Concernant la sélection des logiciels, nous sommes partis d'un raisonnement simple :
Les éditeurs de solution d'analyse et de suivi de référencement sont, par essence, des professionnels du référencement naturel.
Ils se doivent d'être bien positionnés via le moteur de recherche leader[2], il en va de leurs visibilités mais aussi de leurs crédibilités.

Ainsi nous avons exécuté une requête simple non exclusive comportant les mots clés *logiciel d'analyse et de suivi de référencement* dans le moteur de recherche leader en France.

Nous avons ensuite sélectionné les 6 premiers résultats proposés par le moteur de recherche procurant à minima une version d'évaluation à télécharger.

Nous avons diffusé l'ensemble des solutions recueillies au comité de pilotage précédemment formé lors du brainstorming afin que chacun exécute différents tests.

---

[2] En avril 2009, Google enregistré 90% de parts de visites moteurs générées depuis la France (AT Internet Institute, 2009).

57

Puis, nous les avons évaluées collectivement via notre matrice de pondération lors d'une réunion dédiée.

Nous avons ensuite analysé les résultats puis présenté le tout au comité de pilotage et à la Direction pour l'acquisition du logiciel adapté aux besoins de l'entreprise.

## 2  Résultats & Discussion

Lors de la réunion de réflexion, la méthode du brainstorming a permis d'établir les éléments prépondérants suivants :

- Le **c**omparatif entre période : C'est une fonctionnalité indispensable qui permet d'identifier l'évolution et quantifier l'efficacité des actions menées. La plupart des logiciels disposent de cette fonctionnalité mais la précision du réglage des périodes antérieures et la représentation graphique associée peuvent faire la différence.
- L'**i**nteropérabilité : Aujourd'hui, on parle d'une évolution majeure dans le domaine des Web Analytics via la construction de tableaux de bord unifiés pour l'analyse à 360° des investissements marketing en ligne ([WL08]). Mais pour constituer ces tableaux de bord, il est nécessaire de disposer de source de données provenant de solutions comme les logiciels d'analyse et de suivi du référencement dans les moteurs de recherche.

  Dans ce cadre, la planification et le format des rapports sont deux critères primordiaux pour l'agrégation future des données au sein de solutions consolidées.
- Le **t**arif : Le tarif comme pour toute entreprise est un élément important. Mais au-delà du coût, il s'agit d'évaluer les modalités d'accès au logiciel. Une licence est souvent liée à une durée.
- L'**i**nnovation : Dans le domaine du référencement, il y a de nombreuses évolutions liées à la révolution des technologies et des usages issus du Web 2.0. Ainsi, on voit apparaitre de nouveaux concepts permettant de mettre en valeur tout le contenu d'un site web comme la longue traine appliquée au référencement ([And09]). *Le trafic issu de la somme des mots clés de la longue traîne dépasse le total des apports des mots clés les plus performants* ([Ber07]).
- L'**e**rgonomie : L'ergonomie est un terme vaste qui regroupe de nombreux aspects liés à l'appréhension des utilisateurs. *L'ergonomie s'intéresse de manière générale à l'amélioration des conditions de travail, et l'ergonomie du logiciel concentre plus particulièrement* *son attention sur les conditions de travail liées à l'utilisation par l'homme de logiciels interactifs.* ([Bar88]). En ce qui concerne l'entreprise, il est attendu que le logiciel soit intuitif et facile à prendre en main, et qu'il génère des rapports clairs et représentatifs des actions menées par l'équipe e-Marketing.

Les différents éléments proposés ont été regroupés en une approche baptisée CITIE (Comparatif, Interopérabilité, Tarif, Innovation et Ergonomie) par le comite de pilotage.

|  | Yooda See U Rank | SEO Web ranking | SEO Soft | Advanced Web ranking | Agent Web Ranking | SEO Admin-istrator | Trellian SEO Toolkit | WEB CEO |
|---|---|---|---|---|---|---|---|---|
| **Prix** | 2 | 2 | 3 | 2 | 1 | 2 | 2 | 2 |
| **Ergonomie** | | | | | | | | |
| Plateforme | 3 | 2 | 2 | 2 | 1 | 2 | 2 | 3 |
| Rapports | 2 | 3 | 1 | 3 | 2 | 2 | 2 | 2 |
| Moyenne ergonomie | 2,5 | 2,5 | 1,5 | 2,5 | 1,5 | 2 | 2 | 2,5 |
| **Interopérabilité** | | | | | | | | |
| Export | 3 | 1 | 1 | 3 | 2 | 2 | 1 | 2 |
| Interfaçage | 2 | 1 | 2 | 3 | 1 | 1 | 1 | 2 |
| Moyenne interopérabilité | 2,5 | 1 | 1,5 | 3 | 1,5 | 1,5 | 1 | 2 |
| **Innovation** | | | | | | | | |
| Indicateurs | 1 | 1 | 1 | 1 | 1 | 1 | 1 | 2 |
| Fonctionnalités | 3 | 3 | 2 | 3 | 2 | 2 | 2 | 3 |
| Moyenne innovation | 2 | 2 | 1,5 | 2 | 1,5 | 1,5 | 1,5 | 2,5 |
| **Comparaison / Evolution** | 2 | 1 | 2 | 2 | 2 | 1 | 1 | 2 |

**Fig. 2** *Exemple de matrice de pondération « Logiciels d'analyse et de suivi de référencement »*

Les éléments sont étroitement liés aux besoins de l'entreprise et c'est en cela que cette approche ne peut être rendue générique. Dans le cadre d'un nouveau comparatif de logiciel, et même si celui-ci porte sur la même thématique, il sera toujours préférable de renouveler la démarche afin de faire émerger de nouveaux critères du groupe audité. Ceux-ci permet de sensibiliser l'ensemble des acteurs sur la thématique en question et favorise la cohésion quant au choix du logiciel.

A partir des critères qualitatifs exposés par le groupe et des logiciels proposés dans les premiers résultats du moteur de recherche, nous avons complété la matrice de pondération de manière collective (fig. 2).

A partir des résultats, nous avons généré des graphiques de type radar avec marqueurs. Nous avons choisi ce type de graphique pour mettre en valeur une série de résultats pour un logiciel par rapport à un autre, les radars superposés donnent une vue d'ensemble de la pertinence des logiciels.

Pour des raisons de taille, nous nous limitons à la présentation du graphique global, néanmoins nous avions présenté au comité de pilotage des radars graphiques radars par logiciel ainsi que les avantages et inconvénients associés issus des échanges lors de la réunion dédiée à l'évaluation collective des logiciels.

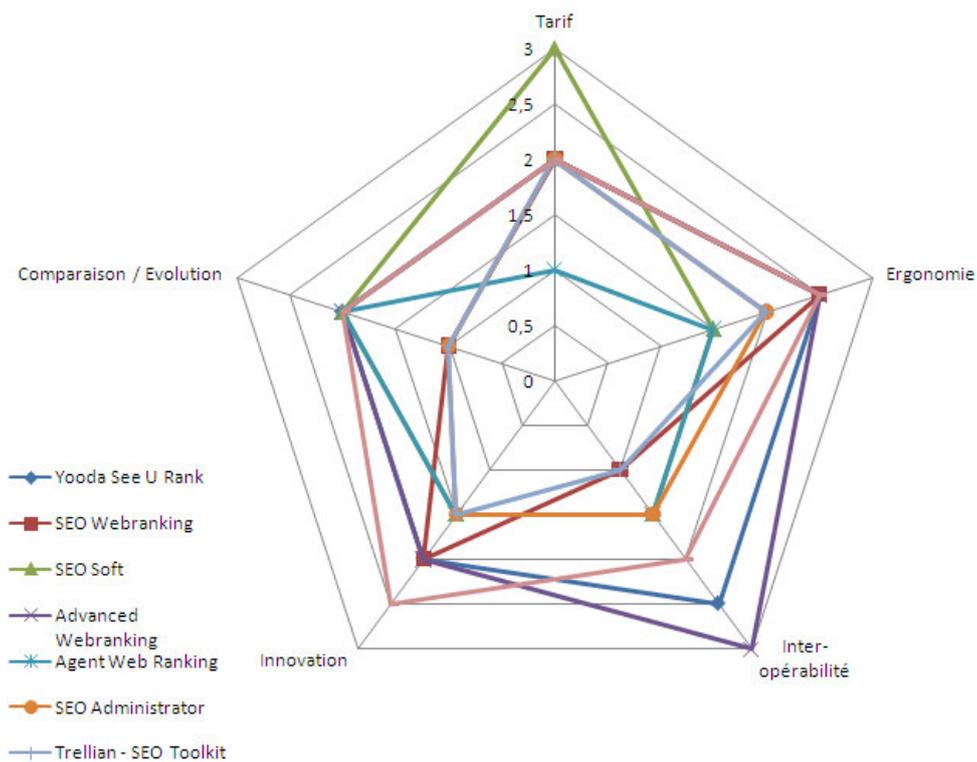

**Fig. 3** *Exemple de graphique global issu de l'évaluation collective*

**Important :** l'évaluation des différents logiciels est étroitement liée aux besoins de l'entreprise et à l'expertise des membres du comité de pilotage qui les ont testés dans le délai imparti. Comme toute évaluation subjective, elle n'engage ni les auteurs de solution, ni les membres du comité de pilotage et l'entreprise qui les embauche.

**Conclusion**

Nous présentons dans cette publication, une démarche qui permet de rapidement selectionner puis évaluer des logiciels pour des entreprises de type PME. Par ailleurs, nous l'avons appliquée à un cas concret matérialisé par la sélection et le choix d'un logiciel d'analyse et de suivi de référencement dans les moteurs de recherche.

Par ailleurs, même si la démarche est relativement récente, il apparait que l'équipe ayant fait le choix du logiciel semble satisfaite du logiciel retenu.

Nous vous invitons à utiliser la démarche et à l'adapter en fonction de vos besoins mais n'oubliez pas que le succès de celle-ci réside dans la mobilisation de l'intelligence collective pour la sélection et le choix d'un logiciel qui répondra aux attentes de votre comité de pilotage.